\documentclass[aps,prb,showpacs,floatfix,amsmath,amssymb,superscriptaddress,reprint]{revtex4-2}
\usepackage[english]{babel}
\usepackage{natbib}
\usepackage{float}
\usepackage{tikz}
\usetikzlibrary{shapes.geometric, arrows.meta, positioning}
\usepackage{bbold}
\usepackage[normalem]{ulem}
\usepackage{color}
\usepackage{graphicx}
\usepackage{epsfig}
\usepackage{setspace}
\usepackage{amsmath}
\usepackage{amssymb}
\usepackage{mathptmx} 
\usepackage{bm}
\usepackage{verbatim}
\usepackage{bibentry}
\usepackage{hyperref}
\usepackage{cancel}
\usepackage{xcolor}
\newcommand{\bea}{\begin{eqnarray}}
\newcommand{\eea}{\end{eqnarray}}

\begin{document}
\title{Mott metal-insulator transition in a modified periodic Anderson model: Insights from entanglement entropy and role of short-range spatial correlations}
\author{Ankur Majumder}%
\author{Sudeshna Sen} 
 \email[Corresponding author: ]{sudeshna@iitism.ac.in}
\affiliation{
  Department of Physics, Indian Institute of Technology, Dhanbad 826004, India.
}
\begin{abstract}
 The Mott transition is a paradigmatic phenomenon where Coulomb interactions between electrons drive a metal-insulator phase transition. It is extensively studied within the Hubbard model, where a quantum critical transition occurs at a finite temperature second-order critical point. This work investigates the Mott transition in a modified periodic Anderson model that
may be viewed as a three-orbital lattice model including an interacting, localized orbital coupled to a delocalized conduction orbital via a second conduction orbital. Within the dynamical mean field theory, this model possesses a strictly zero temperature quantum critical point separating a Fermi liquid and a Mott insulating phase. By employing a simplified version of
the dynamical mean field theory, namely, the two-site or linearized dynamical mean field theory, an analytical estimate is provided for the critical parameter strengths at which the transition occurs at zero temperature. An analytical estimate of the single-site von Neumann entanglement entropy is also provided. This measure can be used as a robust identifier for the phase transition. These calculations are extended to their cluster version to incorporate short-range, spatial correlations and discuss their effects on the transition observed in this model.
\end{abstract}

\maketitle

\section{Introduction}
\label{sec:intro}
The Mott-Hubbard metal-insulator transition (MIT) has been a research focus for decades~\cite{Mott_review}. This is a classic phenomenon observed in strongly correlated electron systems, where the behaviour of the electrons is qualitatively different from the independent electron picture~\cite{mott3,mott1961transition}. The electron-electron repulsions cannot be ignored and are in strong competition with the kinetic energy of the electrons. Typically, unconventional physics arises due to the non-perturbative nature of the electron correlations~\cite{paschen2021quantum}. Indeed, the Mott-Hubbard MIT is a consequence of the non-perturbative nature of the electron correlation effects that become sufficiently larger than the transfer integral between the neighbouring atoms, leading to an energy gap in the charge channel. The Kondo effect, seen in certain metals with magnetic impurities or in quantum dot-like setups, is also a classic consequence of non-perturbative strong correlation effects~\cite{hewson1993kondo}.

The Mott insulator is distinct from a conventional band insulator; the latter is insulating because of its band structure and can thus be understood using the band theory of solids~\cite{ashcroft2022solid}. On the contrary, the Mott transition is a quantum many-body problem requiring methods beyond the one-electron approximation~\cite{Mott_review}. A tenable approach to deal with the complexities posed by a many-body quantum problem is to resort to model Hamiltonians capable of capturing the essential physics of the problem. The single-band Hubbard model famously describes the Mott metal-insulator transition~\cite{mott1949basis,gonzalez1995mott}. The Hubbard model is exactly solvable in two limits, namely, in one dimension and the limit of infinite dimensions ~\cite{lieb1968absence}. The infinite-dimensional limit may be considered the limit of infinite coordination number in spatial coordinates~\cite{vollhardt2022julich}.

Over the last three decades, the dynamical mean field theory (DMFT) framework has been highly insightful in understanding the physics of metal-insulator transitions, heavy fermions, Hund’s metals, etc.~\cite{dmft_rmp}. DMFT is a non-perturbative framework that becomes exact in the limit of infinite coordination number~\cite{metzner_vollhardt_infinite_dim}. The solution of paradigmatic lattice models such as the Hubbard and the periodic Anderson models within the premises of DMFT has enhanced our understanding of strongly correlated electron systems~\cite{Phys_today_DMFT}. Furthermore, DMFT, combined with density functional theory, becomes a robust predictive framework for strongly correlated real materials~\cite{paul2019applications, biermann2006electronic}.

Several studies report on the nature of the metal-insulator quantum phase transition depicted by the Hubbard model~\cite{metzner_vollhardt_infinite_dim,dmft_rmp,bulla1999,bulla2001,Terletska2011,vlad2013}. For example, it is known that the classic Mott MIT, as described by the single-band Hubbard model, exhibits a first-order metal-insulator transition at finite temperatures, and is accompanied by a metal-insulator coexistence regime in the temperature-interaction plane. The first-order line terminates at a second-order critical end-point temperature, $T_c$.~\cite{Terletska2011,vlad2013,Lee_Vojta_2019}. 

In recent works, the Mott MIT has also been interpreted in terms of a new quantity of relevance, namely, the entanglement entropy as derived from the concepts of quantum information theory~\cite{byczuk_quantification_2012,PRX_Tremblay_2020,walsh_local_2019,udagawa_entanglement_2015,larsson_entanglement_2005,larsson_single-site_2006,johannesson_entanglement_2007,gu_entanglement_2004,bellomia_quasilocal_2024,held_physics_2013}. The main idea behind such descriptions start with the calculation of the entanglement entropy using the von Neumann definition~\cite{yang_phase_2000,zanardi_quantum_2002,larsson_entanglement_2005,larsson_single-site_2006,amico2008entanglementRMP}. The von Neumann entropy can serve as an effective quantity in describing systems, especially where the transitions take place without any conventional symmetry-breaking order parameter description. The Mott MIT is thus an extremely pertinent system, and such a description may shed new insights. Recent studies on the Hubbard model in two dimensions have explored the efficacy of local entanglement entropy and mutual information in understanding the order of the Mott transition and its universality class~\cite{PRX_Tremblay_2020,walsh_local_2019,udagawa_entanglement_2015,bera_dynamical_2024,held_physics_2013}.  

In this work, we go beyond the plain vanilla Hubbard model and look into a three-orbital lattice model that may also be considered a bilayer system where a Kondo insulator layer is coupled to a metallic layer~\cite{sen2016quantum}. A more detailed description of the model is provided in Section~\ref{sec:model_formalism}. This model has recently been studied in Refs.~\cite{sen2016quantum} and~\cite{sen2023bilayer} wherein a genuine, quantum critical Mott-Hubbard MIT has been predicted, within the framework of DMFT. Furthermore, unlike the classic Mott transition, the Mott-MIT observed in this system shows quantum critical signatures with an emergent soft-gap power law spectrum at the critical point. One of the goals of this work is to obtain an analytical expression for the critical interaction strength where the transition occurs. Although exact in the infinite dimensions, the solution within the DMFT framework technically requires high-precision numerics, and obtaining analytical estimates is impossible. To achieve our primary goal,  we use a simplified version of the DMFT, namely, the `two-site' or `linearized' DMFT (LDMFT) ~\cite{bulla2000linearized,potthoff_two-site_2001}. Within this framework, we obtain an analytical understanding of the scaling of the quasi-particle weight, $Z$, close to the transition. We also estimate the entanglement content of this model across the transition and show that the entanglement entropy indeed behaves like a robust order parameter for the Mott-Hubbard transition in this model. Currently, there has been no attempt to understand how the nature of the transition in this model gets modified as we include effects beyond DMFT, such as the role of spatial fluctuations, magnetic field, topological effects, etc. Here, we extend the linearized DMFT formalism to incorporate the non-local effects within a minimal cluster framework. We discuss the effects of such minimally included non-local effects on the metal-insulator transition observed in this model and show its effects on the local entanglement entropy.

This paper is organised as follows: in Section~\ref{sec:model_formalism} we review and outline the model and methods, in Section~\ref{sec:results}, we discuss the analytical results obtained and the observations resulting from including non-local effects. Here, we also relate the observations in the context of the entanglement entropy measure of the model within the theoretical frameworks explored in this work. We conclude in Section~\ref{sec:conclusions} and discuss some future directions.



\section{Model and Methods}
\label{sec:model_formalism}
\begin{figure}[htp!]
\includegraphics[clip=,width=\linewidth]{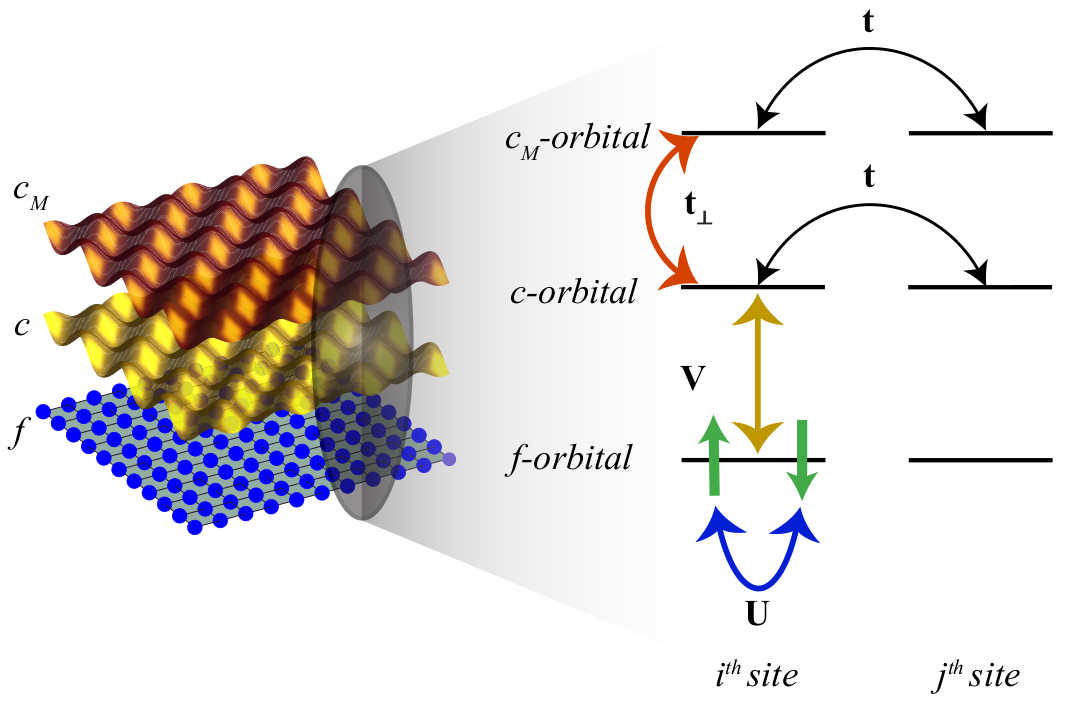}
\caption{A schematic of the model Hamiltonian studied in this work. Each site ($i$) on the lattice consists of three orbitals: a highly localized $f$ orbital (depicted as the blue layer) that hybridizes with a delocalized $c$ orbital (yellow layer) via a hybridization energy $V$. The $c$ orbital is further coupled to a second set of delocalized orbitals ($c_M$, brown layer) via the hopping integral $t_\perp$. The electrons in the delocalized $c$ and $c_M$ orbitals hop around intra-orbitally with hopping energy $t$ between nearest neighbours and are non-interacting. The electrons on the localized $f$ orbitals interact with each other via a repulsive (Hubbard) interaction $U$.}
\label{fig:model}
\end{figure}
 We represent the three orbital lattice model Hamiltonian ($H$), which consists of three parts; $H_f$ represents the interacting, localized $f$ orbital Hamiltonian, $H_c$ ($H_{cM}$) represents the non-interacting, delocalized $c$ ($c_M$) orbitals in the lattice model. The delocalized electrons hop between nearest neighbours (NN) via hopping energy $t$. 
 The $f$ and $c$ orbitals hybridize locally via the hybridization energy $V$, and the two itinerant orbitals couple to each other via $t_\perp$. Combining these components, we can now represent $H$ as,
   $H=H_f+H_U + H_c+H_{cM}+H_{hyb}$, 
where,
\begin{align}
   &H_f + H_U= \epsilon_f\sum_{i\sigma} f^\dag_{i\sigma}f^{\phantom{\dag}}_{i\sigma} + U\sum_i n_{fi\uparrow}n_{fi\downarrow} \\
&H_c + H_{cM}=-t\sum_{ij\in \rm NN, \sigma}\left[(c_{i\sigma}^\dagger
c_{j\sigma}^{\phantom{\dagger}} +\rm{h.c.})+ (c_{Mi\sigma}^\dagger
c_{Mj\sigma}^{\phantom{\dagger}}+\rm{h.c.})\right]\\
&H_{hyb}=V \sum_{i\sigma}( f^\dag_{i\sigma}c_{i\sigma}^{\phantom{\dagger}}+\rm{h.c.}) + t_\perp \sum_{i\sigma}( c^\dag_{i\sigma}c_{Mi\sigma}^{\phantom{\dagger}} + {\rm h.c}  )\,.
\label{eq:ham}
\end{align}
Here, $\sigma=(\uparrow,\downarrow)$ indicates the spin quantum number of the electron occupying the respective orbital. A schematic representation of the above model is shown in Figure~\ref{fig:model}. When $t_\perp=0$, this model reduces to the standard periodic Anderson model, which at half-filling, i.e. $n_f+n_c=2$, renders a particle-hole symmetric Kondo insulating ground state. Thus, this model may also be considered a bilayer Kondo insulator-metal coupled system. In the above model, $\epsilon_f=-U/2$ and on-site energies $\epsilon_c=0$ ($\epsilon_{cM}=0$) refer to the particle-hole symmetric limit. This work focuses on the particle-hole symmetric limit and the conduction electron orbitals arranged on a 2D square lattice. 

Note that the above model reduces to the single-band Hubbard model when the $f$-electrons are dispersive and $V=0$. On the other hand, it reduces to the periodic Anderson model in the limit $t_\perp=0$.
Although the periodic Anderson model does not show any Mott-Hubbard-like metal-insulator transition at half-filling, this model shows a Mott-Hubbard metal-to-insulator transition as the inter-orbital coupling, $t_\perp$, is varied or $U$ is varied, keeping $V$ fixed.  The standard Hubbard model, hosts a finite temperature , classical critical point at $T=T_c$. Nonetheless, a clear quantum critical scaling is still observed in the finite temperature resistivity curves~\cite{Terletska2011}, for $T>>T_c$, presumably originating from an underlying hidden quantum criticality. The crucial difference seen in our model is that, in our case, the $T_c$ is suppressed to $T=0$, and, is not associated with any metal-insulator coexistence region, making it a genuine quantum critical phase transition.~\cite{Lee_Vojta_2019,PhysRevB.104.155114}. Also, unlike the Hubbard model, the metallic solution close to the transition does not show any signature of a pre-formed Mott gap in this model. The reader is referred to a couple of previous studies on this model that demonstrate this transition at zero temperature~\cite{sen2016quantum} and finite temperature~\cite{sen2023bilayer} with variations of these model parameters. In these works, full DMFT calculations were performed using high-precision numerical solvers. Note that the Mott insulating phase can exist only in the range $t_\perp/D\ge1$, where $D$ is the half bandwidth of the conduction electrons. In the following section, we apply the linearized DMFT~\cite{bulla2000linearized} to calculate an approximate expression for the critical interaction strength, $U_c$.

The previous works on this model focused on the infinite-dimensional limit; hence, the DMFT framework was employed in combination with sophisticated impurity solvers. Nonetheless, any numerical simulation always benefits from analytical insights, particularly if such analytical solutions help us obtain scaling relations and expressions describing the behaviour of the quantities that bear signatures of critical transition. In this work, we employ the two-site DMFT approximation introduced by Pothoff and Bulla~\cite{bulla2000linearized,potthoff_two-site_2001}. While it is an extremely crude approximation, it provides qualitatively similar transition parameters to DMFT. However, it is analytically tractable and provides an analytical understanding of the behaviour of the quasi-particle weight. The cluster extension can therefore be straightforwardly explored to understand the possible role of non-local effects on the transition. Finally, we also calculate the entanglement content across the transition within the framework of the two-site cluster DMFT. 

In the following two subsections, we outline the framework and discuss the results in Section~\ref{sec:results}.

\subsection{Linearized Dynamical Mean Field Theory}
In the standard DMFT framework~\cite{dmft_rmp}, the correlated lattice model is self-consistently mapped onto a correlated impurity model, where a single interacting impurity site is embedded into an effective medium represented by a non-interacting bath hybridization function $\Delta(\omega)$. Note that the essence of self-consistent determination of $\Delta(\omega)$ implies that the information of the rest of the correlated sites in the original lattice model has been incorporated into the non-interacting bath for the mapped impurity problem. The locality of the problem ({\it i.e.} mapping of a correlated lattice to a single correlated impurity) essentially involves the assumption that the many-body self-energy is spatially local and has no momentum dependence.  This local self-energy approximation becomes exact in the limit of infinite dimensions, and at any other finite coordination number limit, the DMFT solution is approximate. 

We can first write down the (retarded) lattice $f$-Green's function, $G_f(\omega)$ on the real frequency axis ($\omega$).   
\begin{align}
    G_f(\omega)=\int_{-D}^{D}\frac{\rho_0(\epsilon) d\epsilon}{\omega^+-\epsilon_f-\Sigma_f(\omega)-\Gamma(\epsilon,\omega)},
    \label{eq:Gf_latt}
\end{align}
where $\rho_0(\epsilon)$ denotes the bare density of states of the $c$ and $c_M$ electrons' dispersion; $\Sigma_f(\omega)$ denotes the $f$ electron self-energy embodying the effects of interactions between the electrons. Finally, $\Gamma(\epsilon,\omega)$ contains additional lattice information and is given by, $\Gamma(\epsilon,\omega)=\frac{V^2}{\omega-\epsilon-t_{\perp}^2/(\omega-\epsilon)}$.

The linearized DMFT method~\cite{potthoff2001two} further simplifies the standard DMFT, wherein it is assumed that the bath consists of only one site. In other words, if $H_{\rm imp}$ is the single impurity model to which the lattice model would map, then within linearized DMFT, $H_{\rm imp}=H_{\rm 2-site}$.

This assumption is motivated by the fact that, as one nears the transition, {\it i.e.} $U\to U_c$, the width of the central quasi-particle peak in the local spectral function~\cite{hewson1993kondo}, $A(\omega)=-\mathrm {Im}\frac{1}{\pi}G_{\rm loc}(\omega)$ vanishes. 
Here, $G_{\rm loc}(\omega)$ is the fully interacting impurity Green's function. This central peak also appears in the bath spectral function or $\Delta(\omega)$. 
Thus, within linearized DMFT, one completely ignores the internal structure of the quasi-particle resonance and the high-energy features. The impurity Hamiltonian, $H_{\rm imp}\to H_{\rm 2-site}$ is given by, 

\begin{equation}
    	H_{\rm imp}=\sum _\sigma \left[\epsilon_d d^\dag_{\sigma}d^{\phantom{\dag}}_{\sigma} +\epsilon_{\tilde{c}} \tilde{c}^\dag_{\sigma}\tilde{c}^{\phantom{\dag}}_{\sigma}+\tilde{v} (d^\dag_\sigma \tilde{c}^{\phantom{\dag}}_{\sigma}+\tilde{c}^\dag_\sigma d^{\phantom{\dag}}_{\sigma})\right] + U n_{d\uparrow}n_{d\downarrow}
\label{eq:2-siteham}
\end{equation}
where, $d$ represents the interacting orbital on the impurity, $\tilde{c}$ represents the conduction orbitals, connected via the hybridization energy $\tilde{v}$ in the impurity model. At particle-hole symmetry, the onsite energies, $\epsilon_d=-U/2$ and $\epsilon_{\tilde{c}}=0$.

Using this prescription, we can derive $U_c$ within the linearized dynamical mean-field theory in Section~\ref{susec:LDMFT}. Note that a typical spectral function for the standard Hubbard model consists of a three-peak structure~\cite{byczuk2007kinks} with two Hubbard bands at high energies ($\omega\sim \pm U/2$) and a central quasi-particle resonance. 
Typically, in the conventional Hubbard model, these high and low-energy features are well separated in energy. However, this model is not the case, as shown in Ref.~\cite{sen2016quantum}.  
The local self-energy, $\Sigma_f(\omega)=\Sigma_f^{\rm R}(\omega)+i\Sigma_f^{\rm I}(\omega)$ is also the impurity self-energy by the DMFT construction and is a highly non-trivial quantity to calculate. 
Nonetheless, here we consider only the low energy part of the Fermi liquid self-energy, {\it i.e.}, we approximate, $\Sigma_f(\omega)\sim \Sigma^{\rm R}(0)+\left(1-1/Z\right)\omega+\mathcal{O}(\omega^2)$, 
with $Z=\left[1-\frac{\partial\Sigma^{\rm R}(\omega)}{\partial \omega}\bigg|_{\omega=0}\right]^{-1}$, being the quasi-particle weight that also signifies the width of the central resonance in the $f$ spectral function.

We now substitute this low-energy form of $\Sigma_f(\omega)$ in equation~\ref{eq:Gf_latt} and compare it to the expression for $G_{\rm imp}^d=G_{\rm 2-site}^d$, where the latter can be calculated exactly. Therefore,
\begin{align}
    \lim_{\omega \to 0}G_f(\omega)=\int \frac{\rho_o(\epsilon)\,d\epsilon}{\frac{\omega}{Z}-\frac{V^2\epsilon}{t_{\perp}^2-{\epsilon}^2}}=\int \frac{\rho_o(\epsilon)\,d\epsilon}{\Omega-f(\epsilon)}\textrm{ ,}
    \label{eq:5}
\end{align}
where $f(\epsilon)= \lim_{\omega\to0}\Gamma(\epsilon,\omega) = \frac{V^2\epsilon}{t_{\perp}^2-{\epsilon}^2}$ and $\Omega=\frac{\omega}{Z}$. Since we imposed the condition
that $\frac{\omega}{Z}>>1$ (owing to the fact that $Z \to 0$  as we approach transition from metal side), we can rewrite the equation~\ref{eq:5} as, $\lim_{\omega \to 0}G_f(\omega) \approx \frac{Z}{\omega} + \frac{Z^2}{\omega^2} M_1 + \frac{Z^3}{\omega^3}  M_2 + \mathcal{O}\left( \frac{1}{\omega^4} \right)$, 
where, $M_1=\int_{-D}^{D} \rho_o(\epsilon)f(\epsilon)\,d\epsilon$ and $M_2=\int_{-D}^{D} \rho_o(\epsilon)[f(\epsilon)]^2\,d\epsilon$. 
At particle-hole symmetry, $M_1=0$, so, the lattice green's function for $f$-electrons are given by,
\begin{align}
    \lim_{\omega \to 0}G_f(\omega) \approx \frac{Z}{\omega} + \frac{Z^3}{\omega^3}  M_2 + \mathcal{O}\left( \frac{Z^4}{\omega^4} \right)\textrm{ .}
    \label{eq:low_exp_lat}
\end{align}

The impurity Green's function in the two-site approximation and when $\omega \to 0$ is given by, 
$
    G_{imp}(\omega)=\frac{1}{\frac{\omega}{Z}-\Delta(\omega)}
$
with $\Delta(\omega) \approx \frac{\tilde{v}^2}{\omega}$~\cite{bulla2000linearized,hewson1993kondo}. Thus,
\begin{align}
 \lim_{\omega\to0}G_{imp}(\omega)=\frac{Z}{\omega} + \frac{Z^3}{\omega^3} \frac{\tilde{v}^2}{Z} + \mathcal{O} \left( \frac{Z}{\omega^4}\right)\textrm{ .}
 \label{eq:low_exp_imp}
\end{align}

From comparing the coefficients of $(Z/\omega)^3$ in  $\omega\to0$ expansions of both the local lattice Green's function (eq \ref{eq:low_exp_lat}) and impurity Green's function (eq \ref{eq:low_exp_imp}) in particle-hole symmetry,    
\begin{align}
    \tilde{v}=\sqrt{ZM_2}\textrm{ .}
    \label{eq:1stcon}
\end{align}
This is the self-consistency condition that determines $\Tilde{v}$. When we are away from half-filling, $\epsilon_{\tilde{c}}$ is set by constraining the filling as $n_f=\sum_\sigma\langle n_{f_{i\sigma}}\rangle\overset{!}{=}n_{imp}$. It can be directly calculated from the Green's function as,
$
    n_{f\textrm{ } (imp) }=-\frac{2}{\pi}\int_{-\infty}^0 \Im \left[ G_{f\textrm{ } (imp) }(\omega)\right]d\omega
$.
Thus, the relation reduces,
$    \int_{-\infty}^0 \Im \left[ G_{f}(\omega)\right]d\omega\overset{!}{=}\int_{-\infty}^0 \Im \left[ G_{imp}(\omega)\right]d\omega$. This work is done for the model at the particle-hole symmetry. With the knowledge of $\Tilde{v}$ and $\epsilon_{\Tilde{c}}$, the bath of the two-site impurity Hamiltonian (Ref. equation~\ref{eq:2-siteham}) is completely defined. 

Thus, for given model parameters: $U,V,t_{\perp} \textrm{ and } \rho_o$, $M_2$ can be calculated. Starting with a guess hybridization $\tilde{v}$, the linearized bath is well-defined (assuming the half-filling case). With that information, we solve the $2$-site problems and calculate the impurity of Green's function in Lehmann's representation, finally getting to self-energy. 
This is done using the exact diagonalization method for both the LDMFT and LCDMFT implementations~\cite{lin_exact_1990, lin_exact_1993, Akbar_ED, caffarel_exact_1994,sharma_organization_2015,siro_exact_2012}. Quasi-particle weight is calculated from self energy as $Z=\left[1-\partial_\omega\Sigma^{\rm R}(\omega)\big|_{\omega=0}\right]^{-1}$. After that, the self-consistency condition \ref{eq:1stcon} is imposed to calculate a new hybridization, to be used in the next iteration. Thus, this two-site DMFT solves an Anderson dimer in each iteration until we get a converged value of the $\tilde{v}$. And through this self-consistency scheme, the three-orbital model can be solved in LDMFT formalism.

\subsection{Linearized DMFT: Cluster extension}
\begin{figure}[htp!]
\includegraphics[clip=,width=\linewidth]{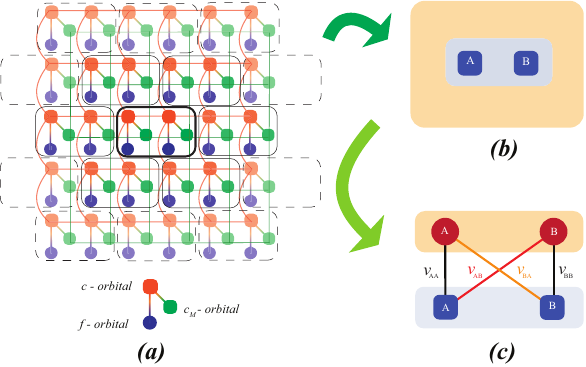}
\caption{Schematic of the cluster framework used in this work. (a) The model Hamiltonian is represented on a 2-dimensional square lattice, with three orbitals on each lattice site, namely, (green) delocalized conduction orbital, $c_M$, locally coupled to a (orange) conduction orbital, $c$. The latter conduction orbital hybridizes with a localized $f$-orbital (blue). This lattice is then mapped within the cellular DMFT framework (CDMFT), considering a pair ($2\times1$) cluster as shown in panel (b). This correlated "impurity" pair (blue shaded sites) is, in principle, embedded in a non-interacting host that needs to be obtained self-consistently within the CDMFT framework. (c) In this work, we further simplify this mapping by considering a linearized-CDMFT (LCDMFT) framework, where the entire host is approximated as two sites (brown shaded sites) coupled to the impurity cluster. Note the hopping scheme used between the impurity cluster and host atoms. ster. Note the hopping scheme used between the impurity cluster and host atoms. }
\label{fig:CDMFT_sch}
\end{figure}
In the standard DMFT approach, the spatial fluctuations are lost due to mapping into an effective single-site quantum impurity problem, which also happens in the case of its linearized simplification. In this section, the investigation is done via a cluster extension of LDMFT with a minimal $2\times1$ cell. We follow the framework in Refs .~\cite {carter_phase_2004, ueda2011theoretical, carter_anisotropic_2004}. 
In standard cluster theory, like cellular DMFT(CDMFT), a cluster of sites is placed into a dynamical mean-field bath, which is determined self-consistently~\cite{quantum_cluster_theories_rmp,senechal201513}. But, in the linearized version, the number of bath sites is approximated to the same number of atoms as in the cluster~\cite{carter_anisotropic_2004,ueda2011theoretical}. In Figure~\ref{fig:CDMFT_sch}, we sketch the outline of the mapping and principle of the linearized cellular-DMFT (LCDMFT) framework.

The local lattice Green's function (in the cluster site basis with sites marked $A$ and $B$), corresponding to the $f$-orbital electrons, is given by (detailed steps of construction are given in appendix \ref{app:lcdmft}), 
\begin{align}
    \bm{G_f}^{-1}(\omega)=\sum_{\vec{k}\in\text{ RBZ}}\begin{pmatrix}
        \omega-\epsilon_{f}-\Gamma_{AA}-\Sigma_{AA}&&-\Gamma_{AB}-\Sigma_{AB}\\
        -\Gamma_{BA}-\Sigma_{BA}&&\omega-\epsilon_{f}-\Gamma_{BB}-\Sigma_{BB}
    \end{pmatrix}
    \label{eq:mainlocal_latt_GF-mat}
\end{align}

where, $\Gamma_{AA}=\Gamma_{BB}=-V^2(\omega-\epsilon_c)(t_{\perp}^2-(\omega-\epsilon_c)^2+\epsilon_k^2)/\eta$, $\Gamma_{AB}=\Gamma_{BA}=e^{ik_x a} V^2(t_{\perp}^2+(\omega-\epsilon_c)^2-\epsilon_k^2)/\eta$, $\eta=(t_{\perp} + \omega - \epsilon_c - \epsilon_k)
(t_{\perp} - \omega + \epsilon_c - \epsilon_k)
(t_{\perp} + \omega - \epsilon_c + \epsilon_k)
(t_{\perp} - \omega + \epsilon_c + \epsilon_k)$.
All the $k$-sums discussed in the context of LCDMFT are actually sums over all $k$-points in the reduced Brillouin zone (RBZ) with proper normalisation. The construction of RBZ for the cluster scheme used here is discussed in Appendix \ref{app:lcdmft}.

Note that, in equation~\eqref{eq:mainlocal_latt_GF-mat}, the $k$-dependence of $\bm{\Sigma}$ is neglected, implying that all non-local effects beyond the cluster length scale have been ignored. Ideally, in CDMFT calculations, one includes the cluster $k$-dependence via a periodization scheme generally implemented on the converged solution~\cite{stanescu2006cellularperiodizn1,cdmftperiodization2,gleis,tremblaycdmft}. The original translational symmetry that was broken due to the clustering can be reinforced by this periodization scheme. Here, this step of periodization is ignored in our approximation while calculating the lattice Green's functions.
Periodization schemes need to be treated with care as they may lead to causality issues in the Green's functions or self-energy, and such implementation is beyond the scope of the current work. Implementations within the framework of DCA will overcome these challenges, and this is left as a future direction.

The local lattice Green's function is now self-consistently mapped to a cluster impurity model given by the Hamiltonian,
\begin{equation}
    H_{imp}=\epsilon_f\sum_{i\sigma}n_{i\sigma}^f + \epsilon_a\sum_{i\sigma}n_{i\sigma}^a +\sum_{ij\sigma}v_{ij}(f_{i\sigma}^\dagger a_{j\sigma}+h.c.)\textrm{ .}
\end{equation}
In this simplification, there are two impurity sites (say site $A$ and $B$) that have the same kinds of bonds as the original lattice $f$-orbital connected to two bath sites given by $a$-operators, $\epsilon_a$ being the on-site energy of bath-sites ; therefore, $(i,j)={A,B}$. The hybridization $v$ that connects the impurity sector to the bath sector (Figure~\ref{fig:CDMFT_sch}(c)) is defined as,
\begin{equation}
    \bm{v}=\begin{pmatrix}
        {f_{A\sigma}}^\dagger&
        {f_{B\sigma}}^\dagger       
    \end{pmatrix}\begin{pmatrix}
        v_{AA} && v_{AB}\\
        v_{BA} && v_{BB}\\
        \end{pmatrix}\begin{pmatrix}
        {a_{A\sigma}}\\
        a_{B\sigma}\\      
    \end{pmatrix}+h.c.\textrm{ .}
    \label{eq:v-mat}
\end{equation}

The impurity Green's function is calculated by integrating out the bath sites as,
\begin{equation}
    \bm{G}_{imp}^{-1}=(\omega  - \epsilon_f )\bm{I}_2- \bm{\Delta}(\omega)-\bm{\Sigma}\textrm{ ,}
\end{equation}
where $\bm{\Delta}=(\omega  - \epsilon_a)^{-1} \bm{v}^\dagger\bm{v}$ is the hybridization function that connects the impurity to the bath sites. At half filling, $\epsilon_a=0$. Impurity Green's function in matrix form is given as,
\begin{equation}
\begin{split}
    \bm{G}_{imp}^{-1}=&\begin{pmatrix}
        \omega-\epsilon_{f}- \Sigma_{AA} && - \Sigma_{AB}\\
        - \Sigma_{BA} && \omega-\epsilon_f - \Sigma_{BB}
    \end{pmatrix}\\
    &-\frac{1}{\omega-\epsilon_a}\begin{pmatrix}
        |v_{AA}|^2 + |v_{BA}|^2 && v_{AA} v_{AB}^* + v_{BB}^* v_{BA}\\
        v_{AA}^* v_{AB}+v_{BB} v_{BA}^*&&|v_{AB}|^2+|v_{BB}|^2
    \end{pmatrix}
\end{split}    
\end{equation}
 
 By following the same prescription as LDMFT (discussed in section \ref{susec:LDMFT}), the self-energy is approximated in coherent regime ($\omega \to 0$) as, $\bm{\Sigma}(\omega)\simeq\bm{\Sigma_0}+(\bm{I}_2-\bm{Z}^{-1})\omega$ where $\bm{Z}=\left[\bm{I}_2-\partial_\omega\bm{\Sigma}(\omega)\big|_{\omega=0}\right]^{-1}=diag(Z_A,Z_B)$. $Z_A$ and $Z_B$ are quasi-particle weights of the $f$ electrons at site $A$ and site $B$, respectively. Thus, in this approximation, the lattice and impurity Green's functions are given by,
 \begin{equation}
     \begin{split}
         \bm{G_f}(\omega)^{(coh)}&=\sum_{\vec{k}\in\text{ RBZ}}\left[\omega \bm{Z}^{-1}- \bm{\zeta} -\bm{\Gamma}(\omega\to0,\bm{\epsilon_k})\right]^{-1}\\
         \bm{G}_{imp}^{(coh)}&=\left[\omega \bm{Z}^{-1}- \bm{\zeta} -\bm{\Delta}(\omega)\right]^{-1}
     \end{split}
 \end{equation}
 where, $\bm{\zeta}=\epsilon_f \bm{I}_2+\bm{\Sigma}_0$. 

 The self-consistency is reached by the condition, $\bm{G}_{f}\overset{!}{=}\bm{G}_{imp}$, that is further modified via a ``molecular orbital basis transformation" discussed in Appendix~\ref{sec:molorb}.
To get an analytical equivalent of $\bm{G}_{f}\overset{!}{=}\bm{G}_{imp}$, the Green functions are expanded in powers of $\omega^{-1}\bm{Z}$ and equating the coefficient of each order. The $\bm{\Delta}(\omega)=\omega^{-1}\bm{v}^\dagger\bm{v}+\mathcal{O}(\omega^{-2})$. The first-order term is trivial. From comparing the coefficient of $\mathcal{O}(\omega^{-2}\bm{Z}^2)$ gives $\sum_{\vec{k}\in\text{ RBZ}}\bm{\Gamma}(\omega\to0,\bm{\epsilon_k})=0$. Similarly, by equating the third order, $\sum_{\vec{k}\in\text{ RBZ}}\left(\bm{\zeta}+\bm{\Gamma}(\omega\to0,\bm{\epsilon_k})\right)\bm{Z}\left(\bm{\zeta}+\bm{\Gamma}(\omega\to0,\bm{\epsilon_k})\right)-\bm{\zeta}\bm{Z}\bm{\zeta}=\bm{v}^\dagger \bm{v}$. Simplifying the last condition by using the result of the previous one gives,
 \begin{equation}
     \sum_{\vec{k}\in\text{ RBZ}}\bm{\Gamma}(\omega\to0,\bm{\epsilon_k})\bm{Z}\bm{\Gamma}(\omega\to0,\bm{\epsilon_k})=\bm{v}^\dagger \bm{v}\textrm{ ,}
     \label{eq:man31}
 \end{equation}
 where $\bm{\Gamma}(\omega\to0,\bm{\epsilon_k})$ is calculated as,
 \begin{equation}
     \bm{\Gamma}(\omega\to0,\bm{\epsilon_k})=\begin{pmatrix}
         0 && e^{ik_x a}f(\epsilon_k)\\
         e^{-ik_x a}f(\epsilon_k) && 0 
     \end{pmatrix}
 \end{equation}
 where $f(\epsilon_k)=\frac{V^2\epsilon_k}{{t_\perp}^2-{\epsilon_k}^2}$. The matrix form of equation \ref{eq:man31} is given by,
 \begin{equation}
     \begin{pmatrix}
         Z_B M_2^\prime && 0\\
         0 && Z_A M_2^\prime
     \end{pmatrix}=
          \begin{pmatrix}
         |v_{AA}|^2+|v_{BA}|^2 && v_{AB}v_{AA}^*+v_{BB}v_{BA}^*\\
        v_{AA}v_{AB}^*+v_{BA}v_{BB}^* && |v_{BB}|^2+|v_{AB}|^2\\
     \end{pmatrix}
     \label{eq:self-con_car}\textrm{ .}
 \end{equation}
 This is the LCDMFT self-consistency equation, where bath parameters (\textit{viz} $\bm{v}$-matrix elements, $\epsilon_a$=0 at half filling) need to be found iteratively. Here, $M_2^\prime=\sum_{\vec{k}\in\text{ RBZ}}\left[f(\epsilon_k)\right]^2$. 
  However, the problem lies in the determination of the bath parameters; there are four equations to solve (considering \textit{SU(2)} spin symmetry). Even using the \textit{U(1)} gauge symmetry ($v_{AA}$ and $v_{BB}$ are taken to be real) the number of unknown reduces to six ($v_{AA}$, $v_{BB}$, $v_{AB}$, $v_{BA}$, $v_{AB}^*$, $v_{BA}^*$), still makes the problem under defined. To resolve this, we need to employ the inversion symmetry within the cluster cell, \textit{i.e.} $A\equiv B$, the equations will remain invariant if $A$ and $B$ are interchanged. We utilize this and employ an orbital basis transformation aimed at diagonalizing the $\bm{v}$-matrix, to arrive at a concise form of the self-consistency condition. This transformation is outlined in detail in Appendix~\ref{sec:molorb}.

 The molecular orbital basis transformation defined via an unitary operator $\bm{U}$ that diagonalize $\bm{v}$ as $\bm{U}^T\bm{v}\bm{U}=\text{\textit{diag} }\left( v_{AA}+v_{AB}, v_{AA}-v_{AB}\right)=\text{\textit{diag} }\left( \Tilde{v}_1, \Tilde{v}_2\right)=\Tilde{\bm{v}}$. Therefore, the self-consistency equation in the molecular orbital basis is given as (the left side of the equation is already diagonal and remains invariant under the transformation), 
\begin{equation}
    \begin{pmatrix}
         \Tilde{Z} M_2^\prime && 0\\
         0 && \Tilde{Z} M_2^\prime
     \end{pmatrix}=\begin{pmatrix}
        |\Tilde{v}_1|^2&&0\\
        0&&|\Tilde{v_2}|^2
    \end{pmatrix}
    \label{eq:self-con:mol}
\end{equation}
where $\Tilde{Z}=Z_A=Z_B$. The quasi-particle weight $(Z)$ in Figure~\ref{fig:z} refers to $Z_{A}$ or $Z_B$ as $A\equiv B$.

\section{Results and Discussion}
\label{sec:results}
\subsection{Linearized Dynamical Mean-field theory}\label{susec:LDMFT}
\begin{figure}[htp!]
\includegraphics[clip=,width=\linewidth]{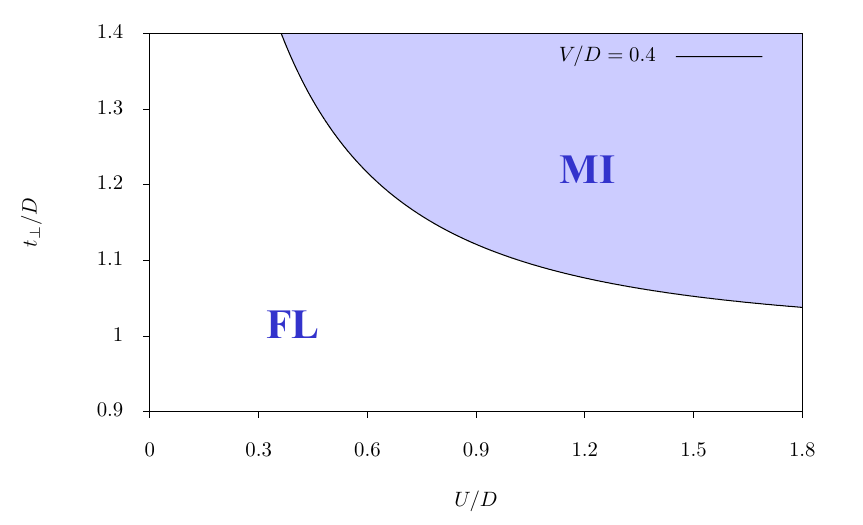}
\caption{\label{fig:phase_diag} Phase diagram of the modified periodic Anderson model for hybridization energy $V/D=0.4$ (connecting $c$ and interacting $f$-orbitals) separating a Fermi liquid (FL) phase from a Mott insulating (MI) phase in the $U-t_\perp$ plane, where $D$ is the half-bandwidth of the conduction orbitals. This phase diagram is evaluated using the expression for $U_c=6\sqrt{M_2}$ (given in equation~\eqref{eq:U_c}), calculated within linearized dynamical mean field theory framework.  Here the information about the lattice, the $c$ electron bare density of states (here chosen 2D square lattice density of state) and the parameter $t_\perp$ (hybridization energy connecting two conduction orbitals $c$ ad $c_M$) enters through $M_2$.}
\end{figure}
In this section, we derive an analytical expression for the critical interaction strength, $U_c$, in this model. We also utilize this analytical expression to get the phase diagram, separating a Mott insulating phase from the Fermi liquid phase, in the $t_\perp-U$ plane for a fixed $V$.
Starting from the DMFT self-consistency, since $G_{f}(\omega)  \overset{!}{=} G_{imp}(\omega)$, 
we must have $\Delta(\omega)=\frac{ZM_2}{\omega}$ by comparing the coherent expansions of impurity and local-lattice Green's functions. 
Furthermore, suppose we now insist on the one-pole structure for the bath hybridization function. In that case, we have $\Delta(\omega) \overset{!}{=} \frac{g_{N+1}}{\omega}$, where, $g_{N+1}$ is the pole wight obtained in the $(N+1)$th DMFT iteration. 
For the particle-hole symmetric case, $Z=\frac{36}{U^2}g_N$~\cite{hewson1993kondo}, such that as $U \to U_c^-$, for a converged solution, $\Delta(\omega)$ is same for successive iterations, and, $g_{N+1} = g_N$. Finally, this gives us
\begin{equation}
    U_c^-=6\sqrt{M_2}\textrm{ .}
    \label{eq:U_c}
\end{equation}
Note that this expression is similar to the conventional Hubbard model~\cite{bulla_linearized_2000} case, but the difference lies in the expression for $M_2$. In Figure~\ref{fig:phase_diag}, we plotted the phase diagram of the model considered, in the $t_\perp-U$ plane for $V/D=0.4$ using the derived expression, $U_c=6\sqrt{M_2}$ for a two-dimensional square lattice. Although derived from a crude approximation, i.e., within the framework of linearized DMFT, the expression benchmarks remarkably well with the expected DMFT results, as was shown previously~\cite{sen2016quantum}. This qualitative agreement between the analytical result derived in this work and the previous numerical calculations obtained with sophisticated DMFT calculations gives us confidence to explore other quantities within this simplified framework. The aim is again not to obtain very precise results but simple analytical estimates that provide us with a preliminary understanding as we increase the level of complexity in the problem. 

\begin{figure}[htp!]
\includegraphics[clip=,width=\linewidth]{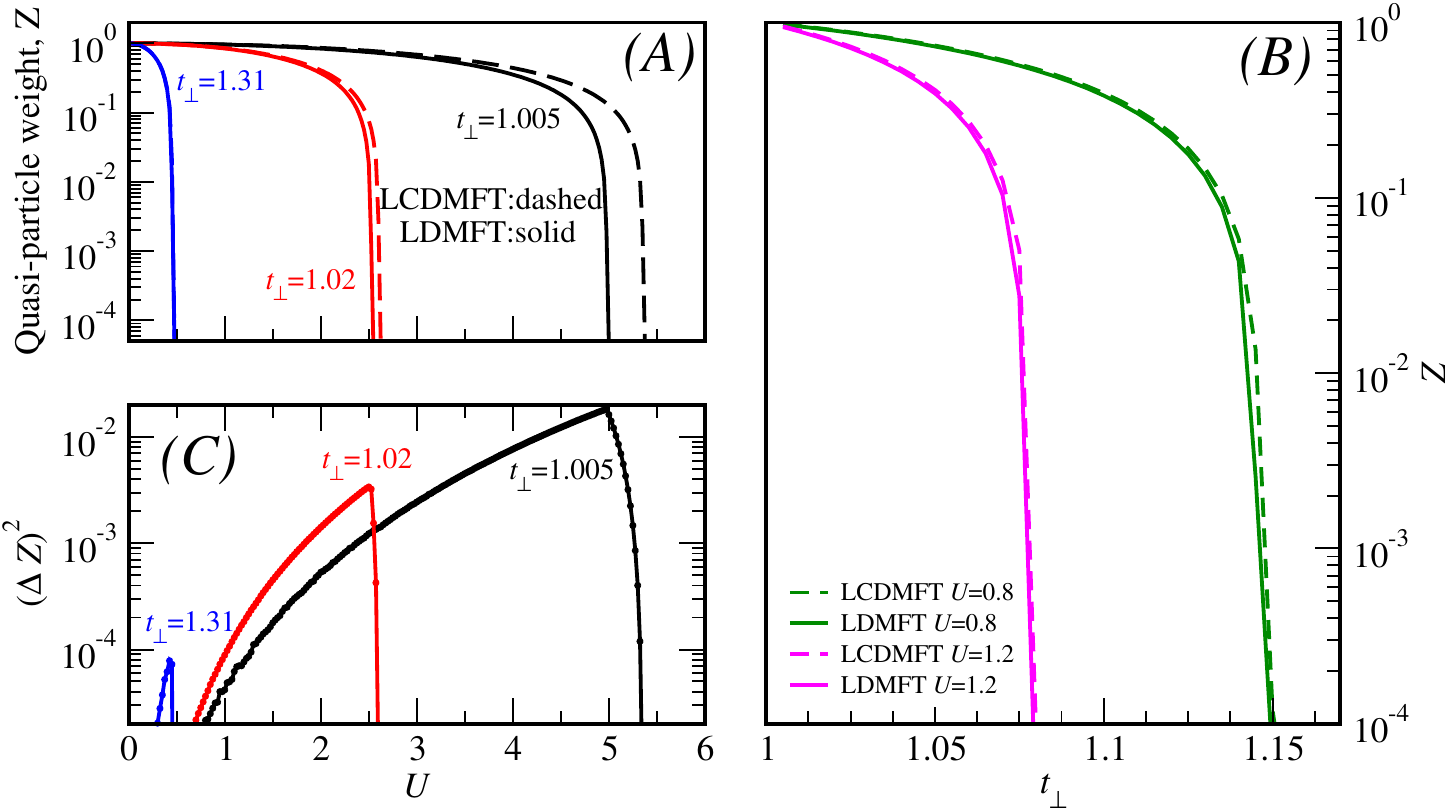}
\caption{The quasi-particle weight, $Z$, calculated within the linearized dynamical mean field theory (LDMFT) and linearized cellular dynamical mean filed theory (LCDMFT) framework, is plotted as a function of the interaction strength, $U$ for $t_\perp/D=1.005,\,1.02\,, 1.31$ in panel (A) and as a function of $t_\perp$ in panel (B), for $U/D=0.8\,, 1.2$. All parameters are scaled by half-bandwidth, $D$. In panel (A), for $t_\perp/D\to 1$, the non-local correlations show a relative increase in the critical interaction strength relative to the local theory. However, as the $t_\perp$ is gradually increased, the local and cluster theories overlap with each other. This could be suggesting that the pair cluster used here is not sufficiently large to capture the non-local correlations due to the increasing $t_\perp$ parameters. The difference is quantified in panel (C), where we plot the quantity $(\Delta Z)^2=(Z_{LDMFT}-Z_{LCDMFT})^2$, where $Z_{LDMFT(LCDMFT)}$ is the $Z$ calculated within the LDMFT (LCDMFT) framework. In  Panel(B), the LDMFT and LCDMFT results overlap as a function of $t_\perp$ for the relatively low $U$ parameters shown.}
\label{fig:z}
\end{figure}
From the Green's function and subsequently the self-energy, $\Sigma(\omega)$ of the 2-site mapped impurity problem~\cite{lange_renormalized_1998}, we can obtain, 
$Z=\frac{1}{1+U^2/36\tilde{v}^2}$. From the self-consistency equation \ref{eq:1stcon}, as we approach the critical point, $U\to U_c$, $Z\to 0$ and $\tilde{v}\to 0$. So, near transition $Z=36\tilde{v}^2/U^2$. Thus, we get, $\tilde{v}=\sqrt{M_2-\frac{U^2}{36}}$. This relation is valid for $U\le U_c$. Putting back the $\tilde{v}$ to equation \ref{eq:1stcon}, we get
    \begin{align}
        Z=1-\frac{U^2}{U_c^2}
    \end{align}
where, $U_c=6 \sqrt{M_2}$, calculated earlier. 

In Figure~\ref{fig:z} we compare the quasi-particle weight $Z$ calculated within LDMFT and LCDMFT as a function of interaction strength $U$ (top left panel A) and as a function of $t_\perp$ (right panel B), where we calculate $Z$ self-consistently using equation~\eqref{eq:1stcon}. The metal-insulator transition observed in this model is indeed captured within this simplified framework. If $U$ is kept constant and $t_\perp$ is varied, such that, $Z\to0$ as $t\to t_{\perp c}$, then one must have, $\frac{U^2}{36 M_2}\to 1$, where, $M_2$ is a positive definite, real function of $t_{\perp c}$~\cite{sen2023bilayer}. 
Of course, the framework does not allow us to capture the transition from the insulating side. We discuss Figure~\ref{fig:z} and the effects of non-local spatial correlations in more detail in Section~\ref{sec:lcdmft}. 

\subsection{Non-local effects within linearized cellular DMFT}
\label{sec:lcdmft}
We now explore the effects of non-local spatial correlations in the most minimal way within the framework of the linearized cellular DMFT (LCDMFT). 
Cellular dynamical mean field theory is a cluster extension of DMFT constructed in real space to include non-local spatial correlations present in a system~\cite{quantum_cluster_theories_rmp,kotliar_cdmft_2008,vlad_cluster_2011, gleis,senechal201513}. There are other methods like the dynamical cluster approximation (DCA) that include the non-local correlation via clustering of $k$-points in momentum space~\cite{quantum_cluster_theories_rmp}. By following the approximations in a two-site or linearized DMFT, the standard CDMFT method can also be simplified to a corresponding simplified version called the linearized CDMFT (LCDMFT).

Before we interpret the result shown in Figure~\ref{fig:z}, we would like to emphasise two unique characteristics of this model. One should note that the coupling parameter, $t_\perp$, plays a two-fold role: (i) it renormalizes the effect of the hybridization energy, $V$, as it introduces additional pathways for the $c$-electrons to decouple from the $f$ electrons, and, therefore, has an effective localizing effect on the $f$-electrons; (ii) secondly, it reduces the effective bandwidth of the $f$-electrons as $V^2/t_\perp^2$~\cite{sen2016quantum}. This leads to a drop in $U_c$ as one increases $t_\perp$. This trend is successfully captured in Figure~\ref{fig:phase_diag}. 

But, now, one may ask, what could interfere with the role of $t_\perp$? Could non-local correlations introduce alternate mechanisms to delocalize the $f$ electrons and mitigate the effects of $t_\perp$ observed in a local theory? Sophisticated cellular DMFT studies on the periodic Anderson model away from half filling highlight the importance of the emergence of RKKY interactions in the vicinity of the quantum phase transition observed in respective studies~\cite{gleis,kotliar_cdmft_2008}. Clearly, this work is beyond the scope of analyzing quantum phase transitions to such great detail. Nonetheless, we ask, what are the minimal effects of non-local correlations on such an analogous model system? In Figure~\ref{fig:z}, we plot and compare the quasi-particle weight, $Z$, calculated within the two schemes, namely, LDMFT (solid lines) and LCDMFT (dashed lines) as a function of (a) $U$, for different $t_\perp$ values fixed (Figure~\ref{fig:z} panel A) and, (b) $t_\perp$, for different $U$ values (Figure~\ref{fig:z} panel B). For a fixed $t_\perp$, when $t_\perp/D\gtrsim 1$, the transition takes place at a higher $U$ than LDMFT, and the non-local effects start showing in the vicinity of the transition. This could indicate either of these two scenarios: (i) $t_\perp\sim D$, is not strong enough to renormalize $V$ and even such a small cluster captures enough short-range spatial correlations that mitigate the localizing effect of $t_\perp$ on the $f$-electrons, so that the $f$ electrons now find it harder to Mott localize, or, (ii) it is an artifact of the size of the cluster and this warrants further investigation. As $t_\perp/D$ is increased, the LCDMFT and the LDMFT results overlap with each other indicating that either, (a) the local dynamical mean field is sufficient to capture the physics, or, (b) the opposite, where, the length scale over which the non-local effects now dominate ranges beyond the size of the cluster considered. However, the individual effects of these scenarios can be captured and distinguished only when we systematically increase the size of the cluster. We quantify this difference in the $Z$ calculated within the two schemes by evaluating the quantity, $(\Delta Z)^2=(Z_{LDMFT}-Z_{LCDMFT})^2$, where $Z_{LDMFT(LCDMFT)}$ is the $Z$ calculated within the LDMFT(LCDMFT) framework and is shown in Figure~\ref{fig:z}(C). In Figure~\ref{fig:z}(B) we fix $U$ at relatively small parameters and calculate $Z$ as a function of $t_\perp$. The results within the two frameworks show no change even for $t_\perp/D\to 1$, and we need to evaluate the $Z$ for a high enough $U$ to observe the non-local changes.

\subsection{Local Entanglement Entropy}
\label{sec:ee}
In this section, we evaluate the local entanglement entropy of the model Hamiltonian discussed in this work. We work within the linearized (or two-site DMFT framework), as our aim is to derive analytical expressions for developing an initial and minimal understanding. Some recent studies also deal with the calculation of quantum entanglement via reduced density matrices to capture the non-local quantum correlations. Theses methods involve diagramatic extensions to include non-local correlations~\cite{PhysRevB.110.075115, bippus2025twositeentanglementtwodimensionalhubbard}. The starting point for such a calculation is quite general, as motivated by several previous studies~\cite{PRX_Tremblay_2020,yeo2019localphillips,RevModPhys.75.1333,byczuk_quantification_2012}. For many-body fermionic Hamiltionians, like the Hubbard model, that cannot be solved exactly, in general, the minimalist approach is to calculate the local entanglement entropy, $S$, involving the entanglement of a single correlated site with the rest of the system. In other words, the entire system, defined by a quantum state $|\psi\rangle$ in the Hilbert space, $\mathcal{H}$ is decomposed into two subsystems, $M$ and $N$. The entanglement entropy will estimate how much of subsystem $M$ is entangled with subsystem $N$ and vice versa, both being equal. Furthermore, the local or single-site entanglement entropy is defined in terms of the von Neumann entropy when the subsystem $M$ consists of just a single site, and the entanglement is calculated using the Fock states of the two subsystems. In particular for lattice models, the system is thus bi-partitioned as one part being one or a cluster of sites, and the remaining environment forms the other part.

\begin{figure}[htp!]
\includegraphics[clip=,width=\linewidth]{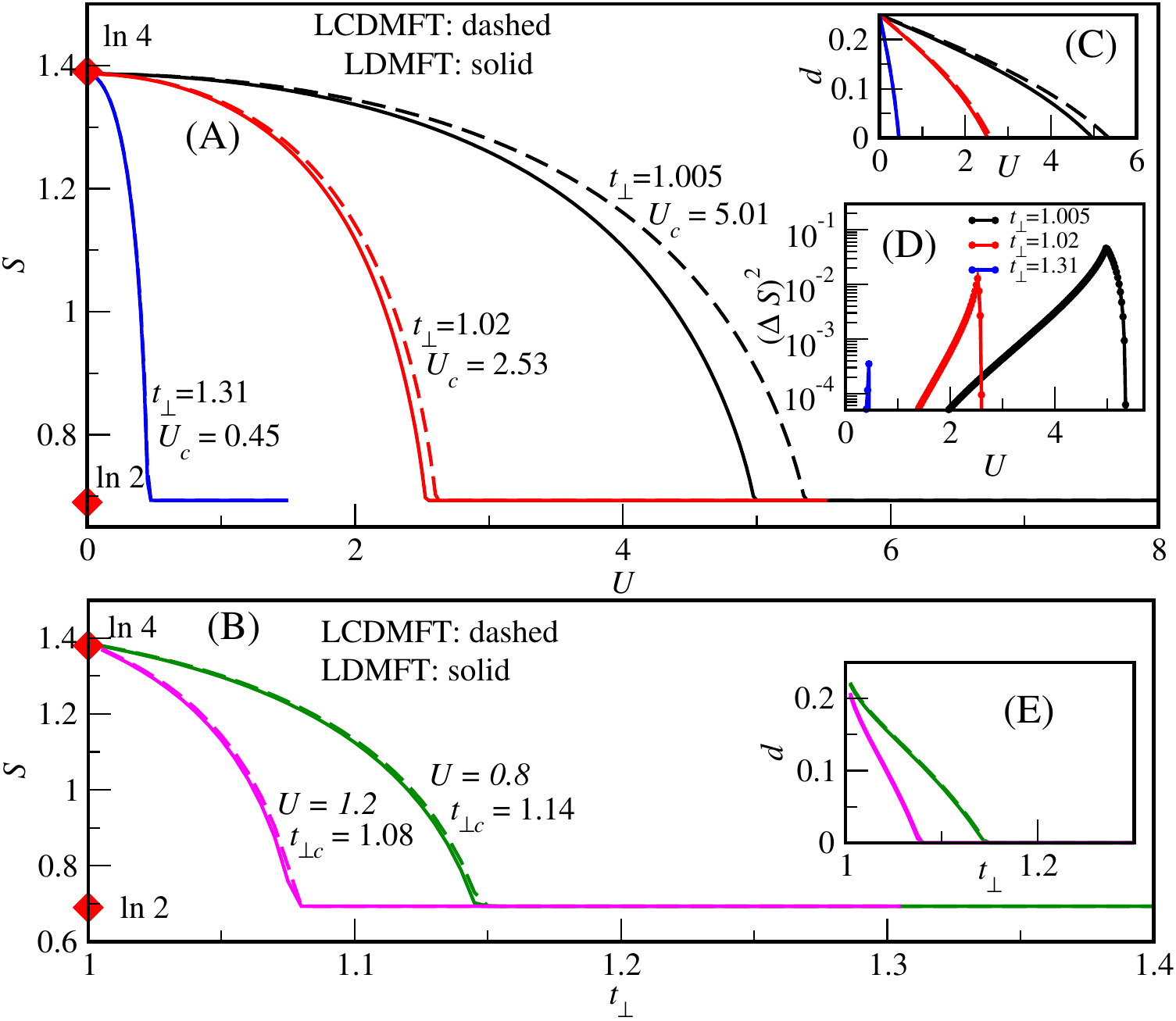}
\caption{(A) The local (von Neumann) entanglement entropy $S$ evaluated using equation~\eqref{eq:entanglement_en}, with the double occupancy, $d$, calculated using linearized DMFT (LDMFT: solid line) and linearized cellular DMFT(LCDMFT: dashed line) are plotted as a function of $U$ for $t_\perp=1.005,\, 1.02,\, 1.31$. 
	(B) The same is plotted as a function of $t_\perp$ for $U=0.8,\, 1.2$. The respective double occupancy plots are shown in the insets (C, E). All parameters are scaled by half-bandwidth, $D$. The free or the non-interacting limit is marked by $S=\ln 4$, and the local moment limit of the model is marked as $S=\ln 2$. (D) $(\Delta S)^2=(S_{LDMFT}-S_{LCDMFT})^2$, where, $S_{LDMFT(LCDMFT)}$ refers to the entanglement entropy evaluated using $d$ calculated within LDMFT(LCDMFT) framework. $(\Delta S)^2$ is plotted as a function of $U$ for the respective $t_\perp$ parameters, highlighting the increasing prominence of non-local effects for decreasing $t_\perp$. Similar observations are shown in Figure~\ref{fig:z}. The $U_c$ values mentioned in (A, B) are obtained using the relation, $U_c=6\sqrt{M_2}$, and the $t_{\perp c}$ values are respectively interpolated.}
\label{fig:entanglement}
\end{figure}
Now, von Neumann Entanglement entropy for this setup is given as $S=-Tr(\hat{\rho}_M\textrm{ ln}\hat{\rho}_M)=-Tr(\hat{\rho}_N\textrm{ ln}\hat{\rho}_N)$ where $\hat{\rho}_M$ and $\hat{\rho}_N$ are reduced density matrices for subsystems $M$ and $N$ respectively. They can be constructed from the density matrix of the whole system as $\hat{\rho}_{M \rm{ / } N}=\textrm{ Tr}_{N \rm{ / } M}\vert \psi_0\rangle\langle\psi_0\vert$ where $\psi_0$ is the ground state of the whole system. The matrix elements of the $\hat{\rho_M}$ is calculated very easily in the basis $\lambda=\{\vert0\rangle,\vert\uparrow\rangle,\vert\downarrow\rangle,\vert d \rangle\}$ (where $\vert d \rangle$ is doubly occupied state) as $\hat{\rho}_M=\rm{diag}\left[\rho_{00},\rho_{\uparrow \uparrow},\rho_{\downarrow \downarrow},\rho_{dd} \right]$ where $\rho_{\lambda \lambda}$ are the probabilities of the particle being in the $\lambda$-state. 
 \begin{equation}
     \begin{split}
     \rho_{dd}&=\langle n_\uparrow n_\downarrow \rangle=d\\
         \rho_{\uparrow \uparrow}&=\langle n_\uparrow(1-n_\downarrow)\rangle=\langle n_\uparrow\rangle - d\\
         \rho_{\downarrow \downarrow}&=\langle n_\downarrow(1-n_\uparrow) \rangle =\langle \ n_\downarrow \rangle -d\\
         \rho_{0 0}&=\langle(1-n_\uparrow)(1-n_\downarrow)\rangle=1-(\langle n_\uparrow \rangle + \langle n_\downarrow \rangle) + d
     \end{split}
 \end{equation} 
We have suppressed the site index as we are considering only a single site as subsystem $M$, and $d$ is the double occupancy of the respective site. At half-filling, $\langle n_\uparrow\rangle=\langle n_\downarrow \rangle=1/2$, the elements of $\hat{\rho}_M$ reduces to $\rho_{dd}=d$, $\rho_{\uparrow \uparrow}=1/2-d$, $\rho_{\downarrow \downarrow}=1/2-d$ and $\rho_{0 0}=d$. The entanglement entropy is then given as,
\begin{align}
    S=-2\left(\frac{1}{2}-d\right)\ln\left(\frac{1}{2}-d\right)-2 d\ln( d )
    \label{eq:entanglement_en}
\end{align}

This is a standard result for Mott-like systems treated within DMFT~\cite{byczuk_quantification_2012}.

 Within the linearized DMFT framework and further using exact diagonalization, the ground state of the two-site impurity problem may then be calculated at half-filling to obtain 
 $
     \vert \psi_0\rangle_{imp}=\frac{1}{\sqrt{2(1+\gamma)}}\left(\gamma \vert \uparrow \rangle_d\vert\downarrow\rangle_{\tilde{c}} + \gamma \vert \downarrow \rangle_d\vert\uparrow\rangle_{\tilde{c}}+\vert d \rangle_d\vert\ 0 \rangle_{\tilde{c}} + \vert 0 \rangle_d\vert d \rangle_{\tilde{c}}\right),
 $
 where $\gamma=\left[8\tilde{v}/(U-\alpha)\right]^2$ and $\alpha=\sqrt{U^2+64\tilde{v}^2}$ (where $\Tilde{v}$ is the hybridization connecting the impurity, with interaction $U$, to the bath site in the 2-site Hamiltonian Ref. equation~\ref{eq:2-siteham})~\cite{lange_renormalized_1998}. Now, we can calculate the $d= \langle \psi_0\vert n_{d_\uparrow} n_{d_\downarrow} \vert \psi_0\rangle=1/2(1+\gamma)$. This can be further simplified to get,
   $d=\frac{(U-\alpha)^2}{4(\alpha^2-U\alpha)}$.
Now, in the perturbative expansion of quasi-particle weight in $\tilde{v}^2/U^2$, if we include up to second order terms in $\Tilde{v}^2/U^2$, we get $Z=36\tilde{v}^2/U^2-1584\tilde{v}^4/U^4$ (in Section~\ref{susec:LDMFT}, only the first term is considered). Putting this back in self-consistency equation \ref{eq:1stcon}, we get a relation of for $\tilde{v}$, given by, 
\begin{align}
   \tilde{v}=\begin{cases}
    \frac{U}{2\sqrt{11}}\left(1-\frac{U^2}{U_c^2}\right) &\quad\text{ $U \to U_c^-$}\\
    0&\quad\text{ $U>U_c$}.
   \end{cases}
\end{align}
When $U\approx U_c$, $\tilde{v}\approx2(1-U/U_c)$, $\alpha\to U+32\tilde{v}^2/U$ and $d\approx 8\tilde{v}^2/U^2$. After putting it all together,
\begin{align}
    d\bigg|_{U\approx U_c}=\begin{cases}
        \frac{4}{11}\left(1-\frac{U}{U_c}\right) &\quad\text{ $U\to U_c^-$}\\
        0 &\quad\text{ $U>U_c$}\textrm{ .}
    \end{cases}
\end{align}
We use this to calculate the entanglement entropy with the equation \ref{eq:entanglement_en} in the limit $U\approx U_c$.

\begin{align}
S\bigg|_{U\approx U_c}=\begin{cases}
-\frac{8}{11}\left(1-\frac{U}{U_c}\right)\textrm{ ln}\left(1-\frac{U}{U_c}\right) &\quad\text{ $U \to U_c^-$}\\
\ln 2 &\quad\text{ $U \ge U_c$}.
\end{cases}
\end{align}
The above is again a standard result for Hubbard-like models~\cite{su_local_2013,byczuk_quantification_2012}. 
Note that in the case of the three-orbital model discussed in this work, $U_c$ is different from the Hubbard model. 
Furthermore, in principle, one can also define a relative entanglement entropy to measure how different two quantum states are. In this case, the quantity $S$ is then subtracted or measured relative to a reference state, which could either be the $U=0$ free state with $S=\ln 4$ or the $U\to\infty$ local moment state with $S=\ln 2$. These limiting values of $S$ also provide us with an unambiguous marker for the transition from a metallic to an insulating state. The transition point is identified as the parameter strength where the value $S$ changes to $\ln 2$. 

In Figure~\ref{fig:entanglement}, we plot the local entanglement entropy, $S$, as a function of $U$ (top panel A) and $t_\perp$ (bottom panel B) evaluated for fixed values of $t_\perp$ and $U$, respectively. The non-interacting, $U=0$ limit is marked by $\ln 4$ entropy, and the strong coupling limit is marked by the local moment limit as discussed previously. Note that in both the schemes, the $S$ is evaluated using equation~\eqref{eq:entanglement_en}, the difference being in the calculation of $d$. In the LCDMFT plots (dashed lines), we choose either site of the two-site cluster ($A$ or $B$, because $A\equiv B$) as the bi-partitioned $M$-subsystem and the rest to be the subsystem-$N$, and use the $d$ obtained via LCDMFT in  equation~\eqref{eq:entanglement_en}. Similarly, the $S$ evaluated within LDMFT uses the $d$ obtained within the LDMFT scheme. Figure~\ref{fig:entanglement}(C) shows the vanishing of the double occupancy, $d$, as the transition is approached, as a function of $U$ for different $t_\perp$, and  Figure~\ref{fig:entanglement}(E) show the same as a function of $t_\perp$ for different $U$. The transition point is marked by the classic signature where $S$ changes to $S=\ln 2$. The transition point obtained from $S$ evaluated within the LDMFT scheme matches excellently well with the analytical estimate of $U_c$ and $t_{\perp c}$. While there is no analytical expression for determining the critical parameters within LCDMFT, we can faithfully rely on the respective $S$ obtained via CDMFT, since at transition, $S=\ln 2$ and remains so thereafter. This highlights the efficacy of local entanglement entropy as a distinct and unambiguous indicator or order parameter for such Mott MITs. 

Figure~\ref{fig:entanglement} shows that even though the LDMFT scheme is a dramatically approximate method, the essential qualitative features are nonetheless captured. The local entanglement entropy measure also captures the trend depicted in Fig~\ref{fig:phase_diag}, namely, with increasing $t_\perp$ ($U$), the $U_c$ ($t_{\perp c}$) decreases (decreases). Finally, in Figure~\ref{fig:entanglement}(D) we plot the quantity, $(\Delta S)^2=(S_{LDMFT}-S_{LCDMFT})^2$ as a function of $U$, for different $t_\perp$ values, where, $S_{LDMFT(LCDMFT)}$ refers to the entanglement entropy evaluated using $d$ calculated within LDMFT(LCDMFT) framework. Similar to the trend observed in Fig~\ref{fig:z}, we find that the non-local effects start laying their impact on the transition as $t_\perp/D\to 1$. 

Studies have shown that kinks and derivatives of entanglement entropy measures may be used to determine the order of a quantum phase transition~\cite{su_local_2013} on a case-by-case basis. It would be interesting to see how the current model behaves as we incorporate such concepts within more accurate schemes incorporating non-local effects. 

\section{Conclusions and Outlook}
\label{sec:conclusions}
In this work, we revisit a simple lattice three-orbital model consisting of a localized, interacting $f$ orbital and two coupled delocalized, non-interacting conduction orbitals, that depicts a Mott MIT at a critical interaction strength, $U$, and coupling, $t_\perp$, between the conduction orbitals. Previous DMFT studies on this model show quantum critical signatures in the form of a power law spectrum at the critical point, displaying a genuine zero temperature quantum critical point, unlike the Mott MIT described by the Hubbard model. Nonetheless, as to any numerical study, the possibilities of exploring analytical limits and framework may shed light on the numerical simulations.

In this work, we build on this question and explore this model Hamiltonian within the simplified LDMFT framework. Using this simplified approach, we obtain an analytical expression relating the critical interaction strength and the lattice parameters, i.e. $U_c = U_c (t_\perp ,V )$. Going beyond the local approximation, we incorporate non-local spatial correlations by implementing this model within a linearized CDMFT framework. We highlight the role of such non-localities on the Mott MIT observed in this model. Furthermore, we quantify the transition using the local entanglement entropy measure using the von Neumann definition. We identify that this quantity can serve as a robust order parameter for Mott-like transitions in many-body systems that evade a conventional symmetry-breaking description. In the non-interacting limit for half-filled lattice systems, $S=\ln4$ (depicting the equal possibility of all the local Hilbert states), and in the Mott insulating phase, $S=\ln2$ (depicting that only singly-occupied states are probable). We calculated the entanglement entropy within both LDMFT and LCDMFT frameworks. The same inference is drawn from this comparison (see fig \ref{fig:entanglement}) as was done from quasi-particle weight. But now the Mott insulating regime is parameterised by local entanglement entropy, and the kink in its variation with $U$ and $t_\perp$ signifies the quantum critical point. Thus, local entanglement entropy serves as a robust quantity that is well-defined in both the Fermi liquid and the Mott insulating phases. 

It should be noted that a key assumption underlying the LDMFT or LCDMFT scheme used here relies on the fact that in the conventional Mott transition, there is a clear separation of energy scales as the transition is approached. However, in this model, a preformed
gap is absent close to the transition, unlike the classic case. While the qualitative features are correctly borne out through this approximate solution, investigations in the context of the critical signatures, like quantum critical scaling, are clearly beyond the scope of such approximate schemes. 

Further developments regarding the fate of the transition in the presence of longer-range non-local spatial correlations need to be investigated. Future studies may include studies of this model within the framework of DCA. However, being described in momentum space, the entanglement entropy measures would need to be accordingly redefined. Cluster extensions within the CDMFT scheme at finite temperatures and utilization of the concepts based on mutual information may also be explored to uncover emergent roles of spatially non-local and thermal effects on the Mott MIT observed in this model.

\section{Acknowledgements}
S.S. acknowledges support from the Science and Engineering Research Board, India (SRG/2022/000495),(MTR/2022/000638), and IIT(ISM) Dhanbad [FRS(175)/2022-2023/PHYSICS]. A.M. acknowledges support from the Prime Minister’s Research Fellowship (PMRF ID- 1602706) from MHRD, India. We acknowledge valuable discussions with Hanna Terletska, Andrew Mitchell and Vladimir Dobrosavljevi\ifmmode \acute{c}\else \'{c}\fi{}.

\appendix
\section{Linearized Cellular Dynamical Mean field theory: Pair cluster treatment of the three orbital model}
\label{app:lcdmft}

In this section, we outline the technical details involved in the pair cluster treatment of the three-orbital model explored in this work. We mainly follow the prescription outlined in the works~\cite{carter_anisotropic_2004,carter_phase_2004} for the Hubbard model. Here, we modify the prescription accordingly for our model. Although most of the equations are similar to those one would obtain for the Hubbard model, we review the necessary steps in this Appendix for completeness.
In standard CDMFT, the original lattice model is mapped to a multi-site impurity model (called the cluster impurity problem), where one cluster cell is embedded in a non-interacting bath that needs to be solved self-consistently. In this simplified version of CDMFT, we approximate our bath size to only one cluster cell. A two-site (both impurity and bath have two sites each) cluster is the simplest unit that can capture the non-local correlation. In this work, a $2\times1$ cluster defined on a 2D-square lattice is chosen. The pair clustering scheme for our model in 2D is pictorially described in Figure~\ref{fig:CDMFT_sch}, where the whole 2D three-orbital lattice is divided into these pair clusters. 
Each cluster has two sites, named $A$ and $B$, placed along the $x$-direction (forming $2\times1$ cells out of a 2D square lattice). Thus, the translational lattice vector modifies to $\vec{T}=n(2a,0)+m(a,a)$ where $n,m$ are integers in this shifted tiling scheme. Within a cluster cell, $(0,0)$ is added to $\vec{T}$ for accessing sub-lattice site-$A$ and $(a,0)$ for sub-lattice site $B$. 
\begin{figure}[htp!]
  \centering
  \includegraphics[clip=,width=\linewidth]{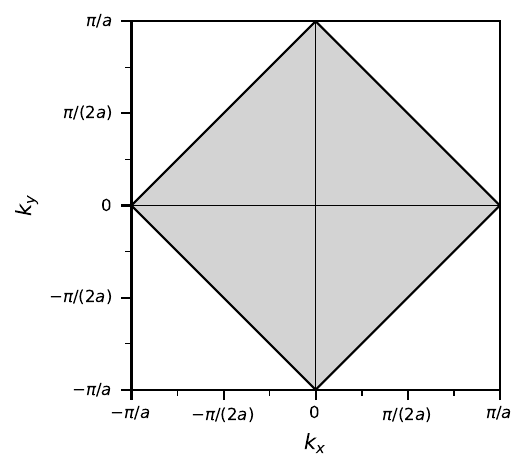}
  \caption{%
    The shaded area represents the reduced Brillouin zone (RBZ) for a 2D shifted‐pair cluster with momentum vector given by $\vec{k}=(k_x,k_y)$, and $a$ is the lattice constant. The shifted tiling scheme is pictorially depicted in Figure~\ref{fig:CDMFT_sch}. 
    \label{fig_RBZ}
  }
\end{figure}
Generally, when working with superlattices, the effective Brillouin zone is reduced depending on the geometric tiling of the clusters. Here, the construction of the RBZ for the pair cluster (specifically, a shifted $2\times 1$ cluster) is discussed. The shifted tiling scheme is used in this work, which involves a half-cell shift in the $y$-direction with respect to the cell below. For this clustering scheme in 2D, the RBZ is defined by the area enclosed by the equation $|k_x|+|k_y|\le\pi/a$, which is shaped like a diamond in momentum space, as shown in Figure~\ref{fig_RBZ}. The Fourier counterpart of the translation lattice vector $\vec{T}$ is the reciprocal lattice vector $\vec{\mathcal{K}}$. A basis for $\vec{\mathcal{K}}$ (\textit{viz,} $\vec{b}_1$ and $ \vec{b}_2$ defined for 2D) can be constructed from the basis of superlattice ($\vec{a}_1\equiv(2a,0)$ and $ \vec{a}_2\equiv(a,a)$) via relation $\vec{b}_i.\vec{a}_i=2\pi\delta_{ij}$. Traditionally, the Bragg's planes as described by $||\vec{k}-\vec{\mathcal{K}}||=||\vec{k}||$ (where $\vec{k}=(k_x,k_y)$ is the momentum vector in 2D) simplifies to $|k_x|+|k_y|\le\pi/a$. It has to be kept in mind that we need to search for the nearest neighbour in reciprocal space, which can only be spanned by $\vec{b}_1$ and $ \vec{b}_2$.

If the tiling scheme is changed, then the shape of RBZ will change. As it turns out, for clustering with no shift in $y$-direction will give a RBZ defined by: $k_y\in [-\pi/a,\pi/a]$ and $k_x\in [-\pi/(2a),\pi/(2a)]$, that gives a rectangular RBZ with same $k$-space area as of shifted case. The rectangular shape of RBZ is intuitively understood as there is super-sizing of the lattice vector along the $x$-direction (by a new lattice parameter $a^\prime=2a$) and the $y$-direction remains the same. Therefore, there is a reduction of the Brillouin zone along the $x$-direction as $k_x\in[-\pi/a^\prime,\pi/a^\prime]$.

The non-interacting Hamiltonian for the three-orbital model written in block form, on the tiled lattice is,  
\begin{equation}
    \bm{\hat{H}_{0,latt}}=\sum_{\vec{k}\in\text{ RBZ}}\begin{pmatrix}
        \bm{f_{k}}^\dagger&
        \bm{c_{k}}^\dagger&
        \bm{c_{Mk}}^\dagger       
    \end{pmatrix}\begin{pmatrix}
        \epsilon_f \bm{I}_{2} && V \bm{I}_{2} &&\bm{0}\\
        V^*\bm{I}_{2} && \bm{\epsilon_k} && t_{\perp}\bm{I}_{2}\\
        \bm{0} && t_{\perp}^*\bm{I}_{2} && \bm{\epsilon_k}
    \end{pmatrix}\begin{pmatrix}
        \bm{f_{k}}\\
        \bm{c_{k}}\\
        \bm{c_{Mk}}       
    \end{pmatrix}\textrm{ ,}
\end{equation}
where $\bm{f_{k}}=\begin{pmatrix}
    f_{k,A} && f_{k,B}
\end{pmatrix}$ represents the annihilation operators of $f$-orbitals in sub-lattice $A$ and $B$. Similarly, $\bm{c_{k}}=\begin{pmatrix}
    c_{k,A} && c_{k,B}
\end{pmatrix}$ and $\bm{c_{Mk}}=\begin{pmatrix}
    {c_M}_{k,A} && {c_M}_{k,B}
\end{pmatrix}$ are the respective $c$ and $c_M$-orbital operators respectively corresponding to sub-lattice $A$ and $B$. $\bm{I}_2$ is identity matrix of order 2. For clear derivation, we have used a notation: when bold symbols are used with a cap on top of them, representing operator description in the full orbital basis, and only the bold letter symbols signify the operator description in the partial basis when the contribution of the other orbitals is separated out. In a pair-clustering system, any bold letter operator is a two-component matrix with sub-lattices $A$ and $B$, which are the components. $\bm{\epsilon_k}$ matrix is given in any one of $\bm{{c_M}_k}$ or $\bm{{c}_k}$  basis as 
\begin{equation}
    \bm{\epsilon_k}=\begin{pmatrix}
        \epsilon_c && e^{ik_x a}\epsilon_k\\
        e^{-ik_x a}\epsilon_k && \epsilon_c
    \end{pmatrix}\textrm{ ,}
    \label{eq:disp}
\end{equation} where, $\epsilon_k=-2t[cos(k_xa)+cos(k_ya)]$; $t$ is the hopping in $c$ and $c_M$-orbitals (Details calculation in Appendix~\ref{app:fourier}). In this notation, we are suppressing the $\sigma$-index for clear representation. At half-filling with particle-hole symmetry, $\epsilon_c=0$ and $\epsilon_{f}=-U/2$. Now, the Green's function in presence of interactions on the $f$-orbital, is given by, $\bm{\hat{G}}^{-1}=\bm{\hat{G}_0}^{-1}-{\bm{\hat{\Sigma}}}=\omega \hat{\bm{I}}-\bm{\hat{H}_{0,latt}}-{\bm{\hat{\Sigma}}}$, where ${\bm{\hat{\Sigma}}}=diag(\bm{\Sigma},\bm{0},\bm{0})$. Here, $\bm{\hat{G}_0}$ is the non-interacting lattice Green's function. As there is no interaction in the $c$ and $c_M$-orbitals, they can be integrated out to get an effective hybridization function for the $f$-orbitals, given by, $\bm{\Gamma}(\omega,\bm{\epsilon_k})=|V|^2\left[\omega \bm{I}_2 - \bm{\epsilon_k}-|t_{\perp}|^2\left(\omega \bm{I}_2 - \bm{\epsilon_k}\right)^{-1}\right]^{-1}$.
This function shows the connection of the $f$-orbitals to $c$-orbital via energy $V$ and indirectly to $c_M$ through $c$ with energy $t_{\perp}$. Thus we reach to the local lattice Green's function as, $\bm{G_f}^{-1}(\omega)=\bm{G}_{0,f}^{-1}(\omega)-{\bm{\Sigma}}(\omega)$.
\begin{equation}
    \bm{G_f}^{-1}(\omega)=\sum_{\vec{k}\in\text{ RBZ}}(\omega - \epsilon_f) \bm{I}_2- \bm{\Gamma}(\omega,\bm{\epsilon_k})-{\bm{\Sigma}}(\omega)
\label{eq:local_latt_GF1}
\end{equation}
The explicit form of this expression is written in the main text equation \ref{eq:mainlocal_latt_GF-mat}.

\section{Molecular orbital basis transformation}
 \label{sec:molorb}
 Based on the above matrix equations, all four elements of Green's functions, self-energy matrices, and $\bm{v}$ are needed, although they are not independent. So we go to a diagonal representation of these matrices via a molecular orbital basis transformation~\cite{ueda2011theoretical}. 
 An orbital transformation is defined for the impurity cluster Hamiltonian via the unitary operation $\bm{U}=e^{-i\frac{\pi}{4}\tau_y}$ ($\tau_y$ is the Pauli matrix in $y$-direction). The transformed impurity Hamiltonian is given by,
\begin{align}    
\Tilde{H}_{imp}=&U^T H_{imp}U=\sum_{i\sigma}\epsilon_f\Tilde{f}_{i\sigma}^\dagger \Tilde{f}_{i\sigma}+\frac{U}{2}\sum_{i\delta}\Tilde{n}^f_{i\uparrow}\Tilde{n}^f_{\delta\downarrow} \nonumber  \\&+\frac{U}{2}\left(\Tilde{f}^\dagger_{1\uparrow}\Tilde{f}_{2\uparrow}\Tilde{f}^\dagger_{1\downarrow}\Tilde{f}_{2\downarrow} + \Tilde{f}^\dagger_{1\uparrow}\Tilde{f}_{2\uparrow}\Tilde{f}^\dagger_{2\downarrow}\Tilde{f}_{1\downarrow} + h.c. \right) \nonumber \\ &+ \sum_{l\sigma}\epsilon_a \Tilde{a}_{i\sigma}^\dagger \Tilde{a}_{l\sigma}+\sum_{l\sigma}\Tilde{v}_{i}\left( \Tilde{a}_{i\sigma}^\dagger \Tilde{f}_{i\sigma} + h.c\right)\textrm{ ,}
\end{align}
where $\bm{\Tilde{c}}=\bm{U}^T\bm{c}$ and $\bm{\Tilde{a}}=\bm{U}^T\bm{a}$. Here, we are using $1$ and $2$ to index sites in this transformed orbital basis instead of $A$ and $B$ that were used to denote the real-space orbitals. The main reason for choosing this kind of transformation is that it can diagonalize the $\bm{v}$ matrix (Ref. equation~\ref{eq:v-mat}), given that there exists an inversion symmetry between the sites. The same transformation can be used to diagonalize the $\bm{G}_{imp/f}$ and $\bm{\Sigma}$ matrices as,

\begin{align}
    \Tilde{\bm{v}}=\bm{U}^T&\bm{v}\bm{U}=\text{\textit{diag} }\left( v_{AA}+v_{AB}, v_{AA}-v_{AB}\right)=\text{\textit{diag} }\left( \Tilde{v}_1, \Tilde{v}_2\right) \textrm{ ,}\nonumber\\
    \Tilde{\bm{G}}_{imp/f}&=\text{\textit{diag} }\left( G_{AA}+G_{AB}, G_{AA}-G_{AB}\right)=\text{\textit{diag} }\left( \Tilde{G}_1, \Tilde{G}_2\right)\textrm{ ,}\\
    \Tilde{\bm{\Sigma}}&=\text{\textit{diag} }\left( \Sigma_{AA}+\Sigma_{AB}, \Sigma_{AA}-\Sigma_{AB}\right)=\text{\textit{diag} }\left( \Tilde{\Sigma}_1, \Tilde{\Sigma}_2\right)\textrm{ .}\nonumber
\end{align}
Therefore, the right side of the self-consistency relation defined in equation \ref{eq:man31} simplifies to \\$\bm{U}^T\bm{v}^\dagger\bm{v}\bm{U}=\bm{U}^T\bm{v}^\dagger\bm{U}\bm{U}^T\bm{v}\bm{U}=\bm{\Tilde{v}}^\dagger\bm{\Tilde{v}}$. Therefore, self-consistency equation in molecular orbital basis is given in the main text equation~\ref{eq:self-con:mol} (the left side of the equation is already diagonal and remains invariant under the transformation). 

\tikzstyle{startstop} = [
  rectangle, rounded corners,
  minimum width=2.8cm, minimum height=1cm,
  text centered, draw=black, fill=red!30
]
\tikzstyle{io} = [
  trapezium, trapezium left angle=70, trapezium right angle=110,
  text width=02.6cm, minimum height=1.4cm,
  draw=black, fill=blue!20,
  align=center, text centered
]
\tikzstyle{process} = [
  rectangle, text width=5.5cm,
  minimum height=1cm,
  text centered, draw=black, fill=orange!30, align=center
]
\tikzstyle{decision} = [
  diamond, aspect=3, text centered, draw=black,
  fill=green!30, inner sep=1pt, text width=5cm, align=center
]
\tikzstyle{arrow} = [
  thick, ->, >=Stealth
]

\begin{figure}
	\centering
	\includegraphics[clip,width=1.0\linewidth]{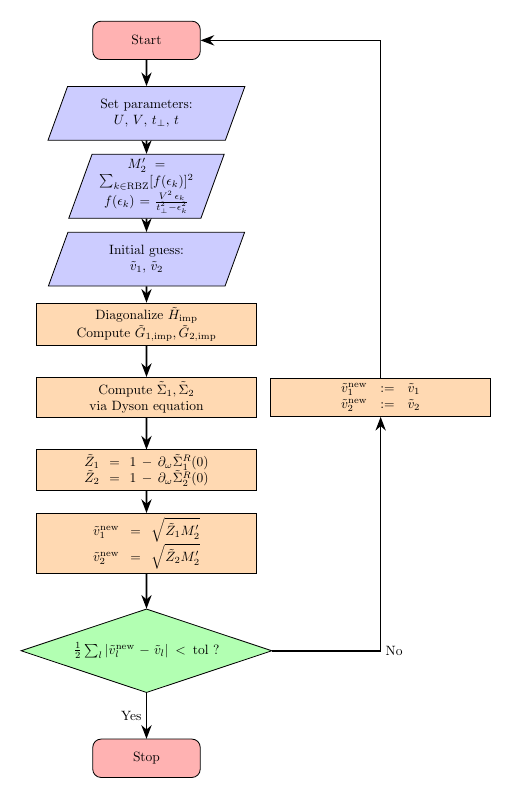}
	\caption{Flowchart of linearized cellular dynamical mean field theory (LCDMFT) implementation for 2D shifted pair cluster for the three orbital model. Here, if inversion symmetry is upheld, $\Tilde{Z}_1\simeq\Tilde{Z}_2=Z$.}
	\label{fig:quasiparticle-flowchart}
\end{figure}


Here, inversion symmetry dictates $Z_{A}=Z_{B}=\Tilde{Z}$. But, if inversion symmetry is there in the cluster, then $\Tilde{v}_1\simeq\Tilde{v}_2$, as implied by the self-consistency equation \ref{eq:self-con:mol}. 
The diagonal terms of the local effective $f$-electron Green's function are calculated as $\bm{\Tilde{G_f}}^{-1}=\bm{U}^T\bm{G_f}^{-1}\bm{U}=\sum_{\vec{k}\in\text{ RBZ}}(\omega - \epsilon_f) \bm{I}_2- \Tilde{\bm{\Gamma}}-{\bm{\Tilde{\Sigma}}}$ where, \\$\Tilde{\bm{\Gamma}}=|V|^2\left[\omega \bm{I}_2 - \bm{\Tilde{\epsilon_k}}-|t_{\perp}|^2\left(\omega \bm{I}_2 - \bm{\Tilde{\epsilon_k}}\right)^{-1}\right]^{-1}$. The only change is in the expression of $\Tilde{\bm{\Gamma}}$ due to this transformation is within $\bm{\Tilde{\epsilon_k}}=\bm{U}^T\bm{\epsilon_k}\bm{U}$. After some simplifications, the diagonal terms of the effective $f$-electron Green's function after molecular-orbital transformation are given by,
\begin{equation}
    \begin{split}
        \Tilde{G_1}=&\sum_{\vec{k}\in\text{ RBZ}}\frac{\omega-\epsilon_f-\Tilde{\Sigma}_2 + \frac{\mathcal{A}}{\mathcal{D}}}{\left(\omega-\epsilon_f-\Tilde{\Sigma}_1 + \frac{\mathcal{B}}{\mathcal{D}}\right)\left(\omega-\epsilon_f-\Tilde{\Sigma}_2 + \frac{\mathcal{A}}{\mathcal{D}}\right)-(\frac{\mathcal{C}}{\mathcal{D}})^2}\textrm{ ,}\\
        \Tilde{G_2}=&\sum_{\vec{k}\in\text{ RBZ}}\frac{\omega-\epsilon_f-\Tilde{\Sigma}_1 + \frac{\mathcal{B}}{D}}{\left(\omega-\epsilon_f-\Tilde{\Sigma}_1 + \frac{\mathcal{B}}{\mathcal{D}}\right)\left(\omega-\epsilon_f-\Tilde{\Sigma}_2 + \frac{\mathcal{A}}{\mathcal{D}}\right)-(\frac{\mathcal{C}}{\mathcal{D}})^2}\textrm{ ,}\\
    \end{split}
    \label{latt_eqn:mol}
\end{equation}
where $\mathcal{D}(\omega)=(t_{\perp}^2-\omega^2)^2-2(t_{\perp}^2+\omega^2)\epsilon_k^2+\epsilon_k^4$, $\mathcal{A}(\omega)=V^2\left[cos(k_x a)\epsilon_k(t_\perp^2+\omega^2-\epsilon^2)+\omega(t_{\perp}^2-\omega^2+\epsilon_k^2)\right]$, $\mathcal{B}(\omega)=V^2\left[-cos(k_x a)\epsilon_k(t_\perp^2+\omega^2-\epsilon_k^2)+\omega(t_{\perp}^2-\omega^2+\epsilon_k^2)\right]$ and $\mathcal{C}(\omega)=V^2sin(k_xa)\epsilon_k(t_{\perp}^2+\omega^2-\epsilon_k^2)$. To calculate the spectral distribution of the original lattice, Green's function must be known in the original basis. So, transformation to our original basis is achieved by an inverse unitary transformation given by $G_{AA}=G_{BB}=(\Tilde{G}_1+\Tilde{G}_2)/2$.

From the local $f$-electrons Green's function expression, both the basis (Ref.~equation.~\ref{latt_eqn:mol} and \ref{eq:mainlocal_latt_GF-mat}), it is evident that calculation of any property at $A$-site (or $1$-site) would have a contribution of from $B$-site (or $2$-site) or vice versa. This shows that the non-local information is getting captured.
In Figure~\ref{fig:quasiparticle-flowchart}, we summarise the entire prescription of LCDMFT as applied to the three orbital model investigated in this work.

\section{Calculation of Dispersion matrix in cluster basis}
\label{app:fourier}
\begin{figure}[htp!]
  \centering
  \includegraphics[clip,width=\linewidth]{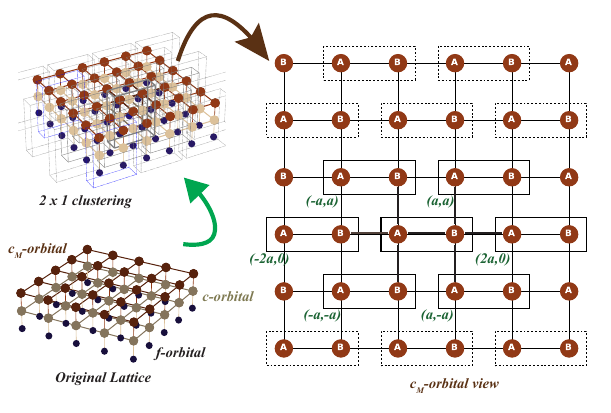}
  \caption{%
    The enlarged figure represents the partial view of the lattice with $c_M$-orbitals (brown shaded orbitals), showing the nearest neighbor bindings of $2\times 1$ cells. The intention behind this is to understand the effect of clustering on the dispersion of the conduction band $c_M$. The same physics shall follow through with the for $c$-orbital.   
    \label{Eps_conc}
  }
\end{figure}
In this section, we review the Fourier transform relations used to relate the superlattice structure~\cite{carter_anisotropic_2004} and the momentum space representation. The $k$-dependence enters through the dispersive orbitals, namely, the $c$ and the $c_M$ orbitals. In the following, we outline the transformation of the non-interacting three-orbital lattice model discussed here.
Fourier transform for the cluster super cell at the position given by $\vec{T}$ will have $k$-vector contribution coming in only within the reduced brillion zone (for our case, it means all the $\vec{k}$ that satisfies $|k_x|+|k_y|\le\pi/a$). It is defined as,
\begin{equation}
c^{\dagger}_{\mathbf{T}, i} = \frac{1}{\sqrt{N}} \sum_{\mathbf{k} \in \text{RBZ}} e^{i \mathbf{k} \cdot \mathbf{T}} \, c^{\dagger}_{\mathbf{k}, i}
\textrm{ ; }c^{\dagger}_{\mathbf{k}, i} = \frac{1}{\sqrt{N}} \sum_{\mathbf{T} \in \text{superlattice}} e^{-i \mathbf{k} \cdot \mathbf{T}} \, c^{\dagger}_{\mathbf{T}, i}\textrm{ .}
\end{equation}
Here, $N$ is the number of $k$-points in the reduced Brillouin zone. 
For our model, there is no dispersion in the interacting $f$-orbital, but there are lattice hopping in the conduction orbitals, $c$ and $c_M$; their hopping is defined in $H_c$ and $H_{c_M}$ terms within the model Hamiltonian (Ref. equation~\ref{eq:ham}). As the terms are independent of each other, we can try to get a general structure of dispersion on the shifted pair cluster basis for one orbital, say $c_M$, and the other one $c$ will have the same structure. In Figure~\ref{Eps_conc}, we have taken a partial view of the $c_M$ orbital after $2\times 1$ clustering of the whole lattice. Considering only nearest neighbour hopping, for $2\times 1$ shifted clustering in 2D, each cluster cell is surrounded by six other cells, which is clear from the figure. The tight-binding Hamiltonian is written as, 
\begin{equation}
\begin{split}
H&_{c_M/c} = -t \sum_{\mathbf{T}} \Big[ \,
 c^{\dagger}_{\mathbf{T},A} c_{\mathbf{T},B} 
+ c^{\dagger}_{\mathbf{T},B} c_{\mathbf{T},A}\\
&+ c^{\dagger}_{\mathbf{T},A} c_{\mathbf{T}+(-2a,0),B}
+ c^{\dagger}_{\mathbf{T},A} c_{\mathbf{T}+(-a,a),B}
 + c^{\dagger}_{\mathbf{T},A} c_{\mathbf{T}+(-a,-a),B}\\
&+ c^{\dagger}_{\mathbf{T},B} c_{\mathbf{T}+(2a,0),A}
+ c^{\dagger}_{\mathbf{T},B} c_{\mathbf{T}+(a,a),A}
+ c^{\dagger}_{\mathbf{T},B} c_{\mathbf{T}+(a,-a),A} 
\Big]\textrm{ .}
\end{split}
\end{equation}
There is a point to note, only the first two terms are hermation conjugate to each other; they represent the intra-bonding within the cluster. Other terms don't have their hermitian conjugate, but they are not left out; they will be accounted for when we add up at the process of every site. This Hamiltonian is transformed to Fourier space with the modification as defined.
\begin{align}
&H_{c_M/c} = -t \,\frac{1}{N} 
  \sum_{\mathbf{T},\,\mathbf{k},\,\mathbf{k}'} 
  e^{\,i(\mathbf{k}' - \mathbf{k})\cdot \mathbf{T}}\,
  \Biggl\{    
     \nonumber \\&c^{\dagger}_{\mathbf{k}',A}\,c_{\mathbf{k},B}\,
    \Bigl[\,1 \;+\; e^{\,i\,2k_x a} \;+\; e^{\,i(\,k_x a + k_y a\,)} \;+\; e^{\,i(\,k_x a - k_y a\,)}\Bigr]
    \;+\; \nonumber\\
    &c^{\dagger}_{\mathbf{k}',B}\,c_{\mathbf{k},A}\,
    \Bigl[\,1 \;+\; e^{-\,i\,2k_x a} \;+\; e^{-\,i(\,k_x a + k_y a\,)} \;+\; e^{-\,i(\,k_x a - k_y a\,)}\Bigr]
  \Biggr\}\,.
\end{align}

After some simplification, we have,
\begin{equation}
H_{c_M/c}= \sum_{\mathbf{k} \in \text{RBZ}} 
\begin{pmatrix}
c^{\dagger}_{\mathbf{k},A} & c^{\dagger}_{\mathbf{k},B}
\end{pmatrix}
\begin{pmatrix}
0 & e^{i k_x a} \epsilon_{\mathbf{k}} \\
e^{-i k_x a} \epsilon_{\mathbf{k}} & 0
\end{pmatrix}
\begin{pmatrix}
c_{\mathbf{k},A} \\
c_{\mathbf{k},B}
\end{pmatrix}.
\end{equation}
This is the structure of $\bm{\epsilon_k}$ given in equation \ref{eq:disp} where $\epsilon_k=-2t[cos(k_xa)+cos(k_y a)]$. Therefore, the full non-interacting Hamiltonian for all the orbitals is written in basis \[
\begin{pmatrix}
  f_{kA} &
  f_{kB} &
  c_{kA}&
  c_{kB}&
  c_{M,A,k}&
  c_{M,B,k}
\end{pmatrix}
\]

\[
\hat{H}_{0,\,\text{latt}} \;=\;
\begin{pmatrix}
  \epsilon_{f} & 0               & V            & 0                & 0                & 0                  \\[6pt]
  0            & \epsilon_{f}    & 0            & V                & 0                & 0                  \\[6pt]
  V^{*}  & 0               & \epsilon_{c} & e^{\,i k_{x} a}\,\epsilon_{k} & t_{\perp}   & 0                  \\[6pt]
  0            & V^{*}     & e^{-\,i k_{x} a}\,\epsilon_{k} & \epsilon_{c}    & 0          & t_{\perp}          \\[6pt]
  0            & 0               & t_{\perp}^{*} & 0                & \epsilon_{c}    & e^{\,i k_{x} a}\,\epsilon_{k} \\[6pt]
  0            & 0               & 0            & t_{\perp}^{*}    & e^{-\,i k_{x} a}\,\epsilon_{k} & \epsilon_{c}
\end{pmatrix}
\]

This formula can be extended to 3D, with a $ 2\times1\times1$ cluster scheme. Only a change to the definition of $\epsilon_k =-2t[cos(k_xa)+cos(k_y a)+cos(k_z a)]$ and the RBZ construction is required.
\vspace{3pt}


\end{document}